\documentclass[reprint,aps,prb,superscriptaddress,showpacs,amssymb,amsmath]{revtex4-1}
\pdfoutput=1
\usepackage{graphicx}
\usepackage{units}
\usepackage{color}
\newcommand{\bra}[1]{\langle{#1}\vert}
\newcommand{\ket}[1]{\vert #1 \rangle}
\newcommand{\bff}[1]{\textbf{#1}}

\newcommand{\Ka}{K_\mathrm{a}}
\newcommand{\Kb}{K_\mathrm{b}}
\newcommand{\sgn}{\mathrm{sgn}}
\newcommand{\buenosaires}{Departamento de F\'{\i}sica, Universidad de Buenos Aires, 
Ciudad Universitaria, Pab.\ I, C1428EHA Buenos Aires, Argentina}
\newcommand{\strasbourg}{IPCMS and NIE, UMR 7504, Universit\'e de Strasbourg -- CNRS, 
23 rue du Loess, BP 43, 67034 Strasbourg Cedex 2, France}
\newcommand{\saclay}{Service de Physique de l'\'{E}tat Condens\'{e}, CNRS URA 2464, 
CEA Saclay, 91191 Gif-sur-Yvette, France}

\begin{document}

\title{Spin-orbit effects in nanowire-based wurtzite semiconductor quantum dots}

\author{Guido A.\ Intronati}
\affiliation{\buenosaires}
\affiliation{\strasbourg}
\affiliation{\saclay}

\author{Pablo I.\ Tamborenea}
\affiliation{\buenosaires}

\author{Dietmar Weinmann}
\author{Rodolfo A.\ Jalabert}
\affiliation{\strasbourg}


\begin{abstract}

We study the effect of the Dresselhaus spin-orbit interaction on the 
electronic states and spin relaxation rates of cylindrical quantum dots 
defined on quantum wires having wurtzite lattice structure.
The linear and cubic contributions of the bulk Dresselhaus spin-orbit 
coupling are taken into account, along with the influence of 
a weak external magnetic field.
The previously found analytic solution for the electronic states of 
cylindrical quantum dots with zincblende lattice structures with 
Rashba interaction is extended to the case of quantum dots with
wurtzite lattices.
For the electronic states in InAs dots, we determine the spin texture
and the effective g-factor, which shows a scaling collapse when plotted 
as a function of an effective renormalized dot-size dependent spin-orbit 
coupling strength.  
The acoustic-phonon-induced spin relaxation rate is calculated and the
transverse piezoelectric potential is shown to be the dominant one.

\end{abstract}

\maketitle


\section{Introduction}
\label{sec:introduction}

Semiconductor nanowires catalytically grown from nanoparticles, also called
nanorods and nanowhiskers, provide a promising platform for spintronic 
devices.
One of their advantages is that small quantum dots can be conveniently 
defined in these nanowires by employing different techniques.
For example, quantum dots can be obtained by varying the composition during 
the growth process \cite{wu-fan-yan,bjo-ohl-sas,lau-gud-wan-lie} in the same 
way that quasi-two-dimensional superlattices are grown using molecular-beam
epitaxy or metalorganic chemical vapour deposition techniques.
This fabrication method produces sharply defined quantum dots with square-well 
confinement in the longitudinal direction with highly controllable lengths.
This type of quantum dot has been thoroughly studied in the last ten years 
and its spin-related properties have attracted much interest.
In particular, the effective g-factor in InAs nanowire-based quantum dots has been 
measured \cite{bjo-fuh-han-lar-fro-sam} for the first few electrons entering 
the dot and a strong dependence on the dot size has been established.
Moreover, studies of spin-orbit interaction and spin relaxation have been 
made in InAs nanowire-based double-quantum dots with two electrons.
\cite{pfu-sho-ens-let,fas-fuh-sam-gol-los}
A less precise but very flexible type of quantum-dot structure can be obtained
by electrostatic confinement given by thin metallic gates deposited 
perpendicularly to the wires, typically made of InAs or InSb. 
\cite{nad-fro-bak-kou,nad-pri-van-zuo-pli-bak-fro-kou,mou-zuo-fro-pli-bak-kou}
The gate voltages, which are generally time-dependent, allow for the 
definition of the quantum dots as well as for the control of charge transport 
and spin manipulation via electric-dipole spin resonance (EDSR). 
\cite{nad-fro-bak-kou,ras-efr,gol-bor-los}

An important aspect of semiconductor nanorods, and also of the quantum 
dots defined in them, is that they often display the wurtzite crystal 
structure even though the constituting material has a zincblende structure 
in the bulk.\cite{kog-kak-yaz-hir-kat}
This structural change has important consequences for the spin properties 
of the dots.
Notably, while in the zincblende semiconductors the leading Dresselhaus 
spin-orbit coupling term is cubic in $\mathbf{k}$, in wurtzite crystals 
a linear term appears.\cite{lew-wil-car-chr}
A realistic theoretical study of the spin-orbit effects in wurtzite quantum 
dots has now become possible due to recent advances in the characterization 
of the band-structure of different wurtzite materials.
While the linear-in-$\mathbf{k}$ Dresselhaus contribution to the conduction 
band energy of wurtzite crystals has been known for some time, 
\cite{lew-wil-car-chr} the cubic term in $\mathbf{k}$ has been obtained 
only recently within the $\mathbf{k}\cdot\mathbf{p}$ approximation 
for different wurtzite semiconductors.\cite{wang-etal,fu-wu}
De and Pryor,\cite{de-pry} on the other hand, calculated the band-structure 
parameters of several binary compounds which normally display the zincblende 
structure in the bulk assuming that they have the wurtzite structure, in order 
to make them available for studies of wurtzite nanowires made of those same 
materials.
These new data pave the way to a realistic study of nanowire-based quantum dots 
with wurtzite structure, which we undertake here.

Disk-shaped quantum dots with the Rashba structural spin-orbit interaction 
have been extensively studied theoretically and have been shown to admit 
an analytical solution for their energy eigenstates, without \cite{bul-sad} 
and with \cite{tsi-loz-gog} an applied perpendicular magnetic field.
This solution, as we show here, can be conveniently extended to wurtzite 
quantum dots having cylindrical symmetry around the crystal c-axis, either 
with flat (``disk") or elongated (``rod") geometry. 
This is the case since the linear wurtzite Dresselhaus coupling 
is mathematically equivalent to the Rashba linear spin-orbit coupling 
characteristic of asymmetric semiconductor quantum wells, and furthermore, 
the newly obtained cubic term of wurtzite admits in the quasi-two-dimensional 
case the same eigenstates as the linear term.
In this work we exploit these similarities in order to give solutions 
of the eigenvalue problem of the wurtzite quasi-two-dimensional structures 
and cylindrical quantum dots.
In a confined geometry, the wurtzite cubic term of the Dresselhaus coupling 
gives rise to an additional linear contribution that reinforces or counteracts 
the bare linear term.
As we will see below, this reinforcement can be actually much bigger than 
the original linear term, opening up an unexplored regime of strong 
``Rashba-like'' spin-orbit coupling in quantum wells and dots.
Also, new possibilities appear when these linear Dresselhaus terms are 
combined with the standard Rashba term due to structural asymmetry. 
Indeed, further reinforcement or cancellations could possibly be achieved 
by tuning the symmetry and dimensions of the structure with the help of an 
external gate voltage.
Thus, flexible schemes of spin-orbit coupling cancellation could be 
implemented leading to very long spin relaxation times in wurtzite 
structures having particular geometric shapes.\cite{har-put-joy}
The Dresselhaus term, enabled by the bulk inversion asymmetry, has been 
shown to yield the dominant coupling mechanism in cases of important 
structural asymmetry, like that of an extrinsic impurities giving rise 
to the impurity band of n-doped semiconductors.\cite{ITWJ}    

The article is organized as follows.
In Section \ref{sec:spinorbitwurtzite} we introduce the effective 
Hamiltonian including the spin-orbit coupling for wurtzite structures.
In Section \ref{sec:quasitwodimensional} we obtain the electronic states 
of quasi-two-dimensional structures considering first only the linear 
spin-orbit coupling (Subsection \ref{sec:linearterm}) and then the full 
Hamiltonian with the cubic spin-orbit term and the Zeeman energy
(Subsection \ref{sec:cubicterm}).
In Section \ref{sec:quantumdots} the solution of the previous Section is 
used to solve the problem of thin cylindrical quantum dots with 
hard-wall confinement potential.
In Subsection \ref{sec:dotenergies} we present the general analytical 
solution of this problem and the energy levels calculated numerically.
In Subsection \ref{sec:spintexture} we explore the spin structure of 
the one-particle eigenstates. 
The experimentally accessible effective g-factor of the quantum 
dots is studied in Subsection \ref{sec:gfactor},
and in Subsection \ref{sec:spinrelaxation} we discuss the spin relaxation 
due to the coupling to phonons.
Section \ref{sec:conclusion} provides concluding remarks.


\section{Intrinsic spin-orbit coupling in wurtzite-based confined geometries}
\label{sec:spinorbitwurtzite}

Within the envelope-function approximation for conduction-band electrons in 
wurtzite semiconductors, the effective quantum-dot Hamiltonian 
\cite{noz-lew,eng-ras-hal} incorporating the linear \cite{lew-wil-car-chr} 
and cubic \cite{wang-etal,fu-wu} Dresselhaus spin-orbit couplings reads
\begin{equation} 
H = H_{\text{0}} + H_{\text{1}} + H_{\text{3}} + H_{\text{Z}},
\label{eq:Htot}
\end{equation}
\begin{equation} 
H_0 = \frac{p^2}{2 \, m^*} + V_{\text{c}}(x,y,z),
\label{eq:Hzero} 
\end{equation}
\begin{equation} 
H_{\text{1}} = \alpha \, \left(k_y \sigma_x - k_x \sigma_y\right),
\label{eq:linear}
\end{equation}
\begin{equation} 
H_{\text{3}} = \gamma \, \left(b k_z^2 - k_x^2 - k_y^2\right) 
                         \left(k_y \sigma_x - k_x \sigma_y\right),
\label{eq:cubic}
\end{equation}
\begin{equation} 
H_{\text{Z}} = \frac{1}{2} g^* \mu_{\text{B}} B \sigma_z,
\label{eq:zeeman}
\end{equation}
where $V_{\text{c}}$ is a nanoscale confinement potential, 
$\boldsymbol{\sigma}$ is the spin operator,
$\alpha$,  $\gamma$, and $b$ are material-dependent parameters, $m^*$ is the 
effective mass and $g^*$ the bulk effective gyromagnetic factor, 
$\mu_{\text{B}}$ is the Bohr magneton, and $B$ is 
an external magnetic field assumed to be applied in the $z$-direction
that coincides with the c-axis of the wurtzite structure.
Here we include the magnetic field only through a Zeeman term since we will 
consider only relatively weak fields whose orbital effects can be safely 
ignored.
In what follows we will consider quasi-two-dimensional and cylindrical 
quantum-dot structures.
Catalytically grown nanorods made out of materials which have the zincblende 
crystal structure in the bulk can adopt either the zincblende or the wurtzite 
structure depending on the size of the nanoparticle seed and other growth 
conditions.
Experimental data allowing to determine $\alpha$, $\gamma$, and $b$
are not yet available, so in our study we will rely on the theoretical 
estimates obtained by De and Pryor.\cite{de-pry}
Motivated by anticipated applications to nanowires, these 
authors calculated all the relevant band-structure parameters assuming 
a wurtzite structure for the semiconductor binary compounds that have 
a zincblende structure in the bulk.
An asymmetry in the $z$-confinement would add a Rashba term, resulting in a 
renormalization of $\alpha$.
In order to give a wider applicability to our results, whenever possible we 
will present them for reasonably large ranges of parameters so that they can 
be adapted to different materials and to parameters newly obtained, 
experimentally or theoretically.


\section{Quasi-two-dimensional systems}
\label{sec:quasitwodimensional}

Before tackling the quantum-dot problem it is useful to consider the 
eigenvalue problem of a quasi-two-dimensional system.
Thus, we choose $V_{\text{c}}=V_{\text{c}}(z)$ which confines the 
electrons only along the $z$-direction, such that $H_0$ can be separated
as $H_0 = H_0^{\text{xy}} + H_0^{\text{z}}$, with an in-plane term
$H_0^{\text{xy}}=(p_x^2+p_y^2)/2 m^*$ and a longitudinal part defined by
$H_0^{\text{z}}=p_z^2/2 m^*+V_{\text{c}}(z)$.


\subsection{Linear term}
\label{sec:linearterm}

If we leave aside for the moment the cubic term $H_{\text{3}}$ and the external 
magnetic field, we are left with a situation mathematically analogous to the classic 
Rashba problem in which the spin-orbit coupling originates from an asymmetric 
extrinsic potential.
Since the total Hamiltonian is separable we can start working with the 
two-dimensional problem in the $(x,y)$ plane given by 
$H_\mathrm{2d} = H_0^{\text{xy}} + H_1$.
Its well-known solution is \cite{ras} 
%
%
\begin {equation}
\zeta_{\mathbf{k}s}(\mathbf{r})= \frac{1}{\sqrt{2A}}
                 e^{i\mathbf{k}\cdot\mathbf{r}}
\left(
      \begin{array}{c}
          s e^{-i \, \left(\varphi_k-\frac{\pi}{2}\right)} \\
          1
      \end {array}
\right),
\label{eq:spinorash}
\end {equation}
%
%
\begin {equation}
  E(k,s) = \frac{\hbar^{2} k^{2}}{2m^{*}}
           - s \alpha k.
\label{eq:energyrashba}
\end {equation}
In these expressions and henceforth, $A$ is the area of the sample, $\mathbf{r}=(x,y)$, 
$\mathbf{k}=(k_x,k_y)$, $k=\sqrt{k_x^2+k_y^2}$,
and $\varphi_k$ is the angle of $\mathbf{k}$ in polar coordinates. 
The spin quantum number $s = \pm 1$ denotes spin-up and spin-down 
eigenstates with respect to the spin quantization axis which lies in 
the xy-plane and is perpendicular to $\mathbf{k}$ with a polar angle 
$\varphi_k -\pi /2$. 
Note that the spin-orbit term in the energy has a minus sign compared to the 
usual Rashba expression, coming from the minus sign used in Eq.\ \eqref{eq:linear}.
The states \eqref{eq:spinorash} are degenerate for given $k$ and $s$.
This plane-wave solution is convenient in most contexts and has the 
advantage that its spin quantization direction is position independent.
However, \eqref{eq:spinorash} does not profit from the fact that the 
$z$-component of the total angular momentum $J_z$ commutes with the 
Hamiltonian and therefore provides a good quantum number, which is an 
extremely useful property when one tackles cylindrically 
symmetric nanostructures.
The common eigenstates of $H_\mathrm{2d}$ and $J_z$ are given by \cite{bul-sad} 
\begin{equation}
 \chi_{m,k,s}(r,\varphi)=
 \begin{pmatrix}
   J_m(kr) \, e^{i m \varphi} \\
   s J_{m+1}(kr) \, e^{i (m+1) \varphi}
 \end{pmatrix}.
\label{eq:spinorashJz}
\end{equation}
The states \eqref{eq:spinorashJz} are degenerate with those of 
\eqref{eq:spinorash} for given $k$ and $s$, and can be 
expressed as superpositions of them.
Note that while the spin of the basis states \eqref{eq:spinorash} lies 
always in the xy-plane, that is not the case for the states 
\eqref{eq:spinorashJz}, which are superpositions of the states
\eqref{eq:spinorash} within degenerate subspaces.
Furthermore, the spin direction in the latter is space-dependent while in 
the former it is not.


\subsection{Cubic term}
\label{sec:cubicterm}

Let us now include the cubic-in-$k$ term of the Hamiltonian, $H_3$, given in 
Eq.\ \eqref{eq:cubic}, and the Zeeman energy, Eq.\ \eqref{eq:zeeman}.
As usual, we work in the envelope-function approximation where the  
Hamiltonian $H$ is expressed by replacing $\mathbf{k}$ by $-i\nabla$.
We adopt cylindrical coordinates ($r,\varphi,z$) and for the in-plane
coordinates we have 
\begin{eqnarray}
H = -\frac{\hbar^2}{2m^*}
        \left(\nabla^2+\frac{\partial^2}{\partial z^2}\right) 
    + V_{\text{c}}(z) 
    + H_1 \nonumber \\
    + \frac{\gamma}{\alpha}\left[b \left(-\frac{\partial^2}{\partial z^2}\right) + \nabla^2 \right]H_1
    + H_Z. 
\label{eq:Htotal}
\end{eqnarray}
where the symbol $\nabla^2$ is used to represent the 
two-dimensional Laplacian.

Assuming that $V_{\text{c}}(z)$ is an infinite potential well of length $L$, the 
proposed solution of the Schr\"odinger equation $H \xi= E \xi$ is
\begin{equation}
  \xi_{nm}(r,\varphi,z)=\psi_{nm}(r,\varphi) \sqrt{\frac{2}{L}}
                        \sin\left(\frac{n\pi z}{L}\right),
  \label{eq:phigrande}
\end{equation}
\begin{equation}
  \psi_{nm}(r,\varphi)=
  \begin{pmatrix}
    u_{nm}(r) \, e^{i m \varphi} \\
    v_{nm}(r) \, e^{i (m+1) \varphi}
  \end{pmatrix},
  \label{eq:psigrande}
\end{equation}
where $u_{nm}(r)$ and $v_{nm}(r)$ are real functions and
$\psi_{nm}(r,\varphi)$ is an eigenstate of $J_z$ with eigenvalue $j_z=m+1/2$.
The corresponding total energy is 
\begin{equation}
E_n^\mathrm{t}=E_n+E_n^z\, ,
\end{equation}
with the radial part $E_n$ and the longitudinal energy
$E_n^z=(\hbar^2/2m^*)(n\pi/L)^2$ that is due to the confinement in the $z$-direction.
Plugging \eqref{eq:phigrande} into the Schr\"odinger equation we obtain 
for $u_{nm}$ and $v_{nm}$ the equations
\begin{widetext}
\begin{subequations}
\label{eq:uv}
\begin{align}
    \left(-\nabla_m^2+h\right) u_{nm}(\rho)
    +\left(\alpha_n'+\gamma' \nabla_m^2\right)
    \left(\frac{m+1}{\rho}+\frac{\partial}{\partial\rho}\right)v_{nm}(\rho)
    =\varepsilon_n u_{nm}(\rho)
\label{eq:u} \\
    \left(-\nabla_{m+1}^2-h\right) v_{nm}(\rho)
    +\left(\alpha_n'+\gamma'\nabla_{m+1}^2\right) 
    \left(\frac{m}{\rho}-\frac{\partial}{\partial\rho}\right)u_{nm}(\rho)
    =\varepsilon_n v_{nm}(\rho)
\label{eq:v}
\end{align}
\end{subequations}
\end{widetext}
where
\begin{equation}
  \nabla_m^2 \equiv \frac{1}{  \rho}\frac{\partial}{\partial\rho}
                          + \frac{\partial^2}{\partial\rho^2}
                          - \frac{m^2}{\rho^2}. 
\end{equation}
In Eqs.\ \eqref{eq:uv} we have introduced $R$, a parameter 
to be defined in the quantum dot context, and 
$u_{\text{\tiny E}}=\hbar^2/2m^*R^2$, as units 
of length and energy, respectively.
This allows us to define the dimensionless parameters $\rho = r/R$, $K=kR$, 
$\gamma'=\gamma/u_{\text{\tiny E}}R^3$, and 
$h = g\mu_{\text{B}}B/2u_{\text{\tiny E}}$.
The dependence on the (``longitudinal") quantum number $n$ has been 
incorporated to the in-plane problem via the redefinition
of the coupling constant $\alpha$ written in the dimensionless form
\begin{equation}
  \alpha_n'=\left[\alpha+\gamma b \left(\frac{n\pi}{L}\right)^2\right]/
            u_{\text{\tiny E}} R\, ,
  \label{eq:alpha_redef}
\end{equation}
and the in-plane dimensionless energy is given by $\varepsilon_n=E_n/u_{\text{\tiny E}}$. 

To solve Eqs.\ \eqref{eq:uv} we make the ansatz
\begin{eqnarray}
u_{nm}(\rho)=J_m(K\rho), \hspace*{0.5cm}
v_{nm}(\rho)=d_n J_{m+1}(K\rho).
\label{eq:enterbessel}
\end{eqnarray}
Using the well-known properties of the Bessel functions \cite{abramowitz-stegun}
\begin{subequations}
\begin{eqnarray}
\left(\frac{m}{\rho}-\frac{\partial}{\partial\rho}\right)J_m(K\rho)
=K J_{m+1}(K\rho), \\
\left(\frac{m+1}{\rho}+\frac{\partial}{\partial\rho}\right)J_{m+1}(K\rho)
=K J_m(K\rho),
\label{eq:raising}
\end{eqnarray}
\end{subequations}
one obtains from \eqref{eq:uv} the eigenvalue equation 
\begin{equation}
  \begin{pmatrix}
    K^2 +h - \varepsilon_n         &  \alpha_n'K- \gamma' K^3 \\
    \alpha_n' K-\gamma' K^3  &  K^2 -h - \varepsilon_n
  \end{pmatrix}
  \begin{pmatrix}
      1 \\
    d_n
  \end{pmatrix}
= 0 \, ,
\end{equation}
whose solutions are
\begin{equation}
  \varepsilon_{n\pm}=K^2\pm \sqrt{K^2\left(\alpha_n'-\gamma'K^2\right)^2+h^2}\, .
\label{eq:solutionwell}
\end{equation}
Then, the total energy is given by
\begin{equation}
  E_{n\pm}^\mathrm{t}=(\varepsilon_{n\pm}+\varepsilon_n^z)u_{\text{\tiny E}}\, ,
  \label{eq:solutionwell_with_z}
\end{equation}
with $\varepsilon_n^z=E_n^z/u_{\text{\tiny E}}=(n\pi R/L)^2$,
and the corresponding wave functions are 
\begin{equation}
  \psi_{Knm}(\rho,\varphi)=
  \begin{pmatrix}
    J_m(K\rho) \, e^{i m \varphi} \\
  d_{n\pm}\,  J_{m+1}(K\rho) \, e^{i (m+1) \varphi}
  \end{pmatrix},
  \label{eq:psigrandeJ}
\end{equation}
with
\begin{equation}
d_{n\pm} =  \frac{\varepsilon_{n\pm} - K^2 - h }
                 {\alpha_n' K - \gamma' K^3} \,.
\label{eq:d_n}
\end{equation}
The obtained solution, Eqs.\ \eqref{eq:solutionwell} and \eqref{eq:psigrandeJ},
reduces to the one of the linear Hamiltonian analyzed in 
Sec.\ \ref{sec:linearterm}, given by Eqs.\ \eqref{eq:energyrashba} 
and \eqref{eq:spinorashJz}, when the cubic term and the Zeeman energy 
are neglected.
As in the linear Rashba-like problem, there are two possible energies 
$\varepsilon_{n\pm}$ for a given value of $K$.
The energies $\varepsilon_{n\pm}$ can be expressed as a function of $K^2$ 
and are thus independent of the sign of $K$. 
Because of the (anti-)symmetry of the Bessel functions with respect to a 
change of sign in the argument, the wave functions corresponding 
to $\pm K$ are not independent. 
We therefore keep only positive values of $K$. 

In the presence of a magnetic field, and for in-plane energies 
$\varepsilon_n$ close to zero, Eq.\ \eqref{eq:solutionwell} has solutions with 
imaginary $K=i\kappa$. Since $J_{m}(i \kappa \rho) = i^m I_m(\kappa \rho)$, 
where $I_m$ is the modified Bessel function of the first kind of order $m$, 
the corresponding wave functions grow exponentially with increasing $\rho$ 
and are thus not normalizable in an infinitely large system. Such solutions 
are therefore discarded in the context of two-dimensional systems, but they 
will become relevant for the case of quantum dots discussed in 
Sec.\ \ref{sec:quantumdots}.

\begin{figure}
\centerline{\includegraphics[width=\linewidth]{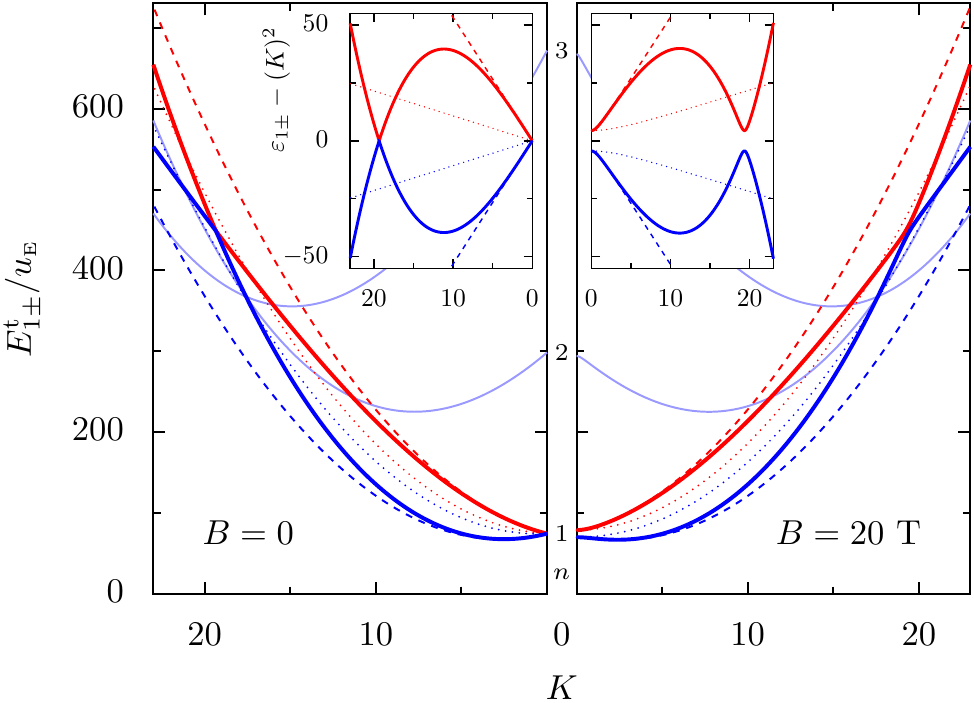}}
\caption{
Energy dispersion [Eq.\ \eqref{eq:solutionwell_with_z}] with (right) and
without magnetic field (left).
For subband $n=1$, three cases are considered: 
thick solid lines correspond to the full SOC, dashed lines to an 
intermediate case with no cubic-in-$\mathbf{k}$ SOC, 
but with the $\alpha$ parameter renormalized by $\gamma$ \eqref{eq:alpha_redef},
and dotted lines to the bare Rashba-like interaction, linear in $\mathbf{k}$.
Blue (red) lines correspond to $\varepsilon_{n-}$ ($\varepsilon_{n+}$).
The thinner solid lines are the lower branches of subbands $n=2$ and 3.
{\it Inset}: Same three cases of SOC dispersion relation without the parabolic 
contribution for $n=1$. 
The Zeeman effect and an avoided crossing are more clearly
distinguished on this energy scale.}
\label{fig:energyDisp}
\end{figure}
In Fig.\ \ref{fig:energyDisp} we present (solid lines) the dispersion 
relation \eqref{eq:solutionwell_with_z} for InAs with the parameters 
suggested in Ref.~[\onlinecite{de-pry}] from band-structure calculations 
(we label them with an index r)
\begin{equation}
  \alpha_{\mathrm{r}} = \unit[0.571]{e\text{V\AA}} \;\;\;\;\; 
  \gamma_{\mathrm{r}} = \unit[571.8]{e\text{V}\text{\AA}^3} ,
  \nonumber
\end{equation} 
$b=4$ and an effective mass $m^*=0.026 \, m_e$.\cite{bjo-fuh-han-lar-fro-sam}
The two energy branches are plotted for $B=\unit[0]{\text{T}}$ (left) 
and $B=\unit[20]{\text{T}}$ (right).
(In this figure we consider a large value of the magnetic field 
with the sole purpose of illustrating more clearly its effects 
on the energy levels.)
We also show the effect of suppressing the cubic term, but keeping 
the contribution of $\gamma$ in Eq.\ \eqref{eq:alpha_redef} on the linear 
term (dashed line), as well as the usual Rashba-like case obtained for 
$\gamma=0$ (dotted line). 
Blue (red) lines correspond to $\varepsilon_{n-}$ ($\varepsilon_{n+}$).
Thick lines correspond to $n=1$, as indicated between the two panels.
Also shown are the curves of $\varepsilon_{2-}$ and $\varepsilon_{3-}$
including the linear and cubic spin-orbit contributions (thin lines). 
For $n=1$ and $B=\unit[0]{\text{T}}$ there is a crossing of the two branches
at $K=\sqrt{\alpha'/\gamma'}$ ($K = 19.34$ in our plot).
An analogous feature present in bulk wurtzite semiconductors has been discussed 
in the literature as a possible opportunity
to implement long-lived spin qubits.\cite{wan-wu-chi-lo}
The crossing becomes avoided for finite $B$, although the level splitting
can hardly be seen on the right panel of Fig.\ \ref{fig:energyDisp}. 
For this reason we plot in the inset the energies subtracting the trivial
parabolic contribution.
This allows for a smaller energy range such that one can clearly observe 
the Zeeman splitting at $K=0$ and the avoided crossing.

In Fig.\ \ref{fig:energyDisp}, the thin solid lines are the lower branches 
of subbands $n=2$ and 3.
Even though they lie at sufficiently high energies so as not to affect
our further analysis, which concentrates on low energies, we note that 
they could become relevant if the region of the avoided crossing mentioned
above is explored.
Also, we point out a potentially interesting degeneracy point of all
the lower branches of the different subbands, which happens at
$K=1/b\gamma'$ ($K = 17.58$ in our plot), where the curves become
independent of $n$.
This massive degeneracy is due to the renormalized linear spin-orbit term.
Although this feature may be physically relevant, we mention that higher 
values of $n$ correspond to higher $k_z$ and eventually the energies of 
Eq.\ \eqref{eq:solutionwell_with_z} obtained in third-order perturbation 
theory in wave-vector cease to be reliable.


\section{Quantum dots}
\label{sec:quantumdots}

\subsection{Effect of spin-orbit coupling on the energy levels}
\label{sec:dotenergies}

We now consider cylindrical quantum-dots with hard-wall quantum confinement 
having radius $R$ and length $L$.
The discrete eigenenergies and states of this problem will be obtained from the
quantum-well solutions found in the previous Section.
In order to get the energetically lowest states, we keep only the lowest
subband, $n=1$, and omit the subindex $n$ from now on.
In all cases we work with $k$ low enough to stay in the regime of validity of the 
expansion of the effective SOC Hamiltonian up to 3\textsuperscript{rd} order in $k$.

The eigenstates of the disk-shaped quantum dot have to satisfy 
the circular boundary condition (the hard-wall confinement forces a
zero of the wave function at the dot boundary). This can be achieved at 
particular values of the in-plane energy $\varepsilon$ for  
linear combinations $\Psi_m=c_\mathrm{a} \psi_{\Ka m}+c_\mathrm{b}\psi_{\Kb m}$ 
of two degenerate eigenstates of the quantum-well problem. 
Those quantized energies are then the eigenenergies of the quantum dot. 
In the general case including a finite magnetic field, there are 
three energy ranges (see Fig.\ \ref{fig:energyDisp}) with different 
situations:
i)  energies in the low ``belly" of the $\varepsilon_{-}$ branch, $\varepsilon<-|h|$;
ii) energies above the energy gap caused at $K=0$ by the Zeeman splitting, 
$\varepsilon>|h|$;
iii) energies in the Zeeman gap, $-|h|<\varepsilon<|h|$. 
We now consider these three cases separately.

\noindent
\textit{Case i)} $\varepsilon<-|h|$: 
two real values of $K$, noted $\Ka$ and $\Kb$, associated to the 
$\varepsilon_{-}$ branch (in the ``belly" region) are involved 
in the dot solution.
The in-plane wave function is thus written as 
\begin{widetext}
\begin{equation}\label{eq:state_casei}
\Psi_m(\rho,\varphi) = c_\mathrm{a} \begin{pmatrix}
                     J_{m}(\Ka \rho)e^{im\varphi}\\ 
                     d_{-}(\Ka) J_{m+1}(\Ka \rho)e^{i(m+1)\varphi}
                   \end{pmatrix} + 
               c_\mathrm{b} \begin{pmatrix}
                     J_{m}(\Kb \rho)e^{im\varphi} \\  
                     d_{-}(\Kb) J_{m+1}(\Kb \rho)e^{i(m+1)\varphi}
                   \end{pmatrix},
\end{equation}
with the hard-wall boundary condition $\Psi_m(\rho=1,\varphi) = 0$.
A non-trivial solution $(\Ka, \Kb )$ will be given by the condition
\begin{equation}
  J_{m}(\Ka) \, d_-(\Kb) J_{m+1}(\Kb)-J_{m}(\Kb) \, d_-(\Ka) J_{m+1}(\Ka) = 0 \,.
\label{eq:quantization_casei}
\end{equation}
\noindent
\textit{Case ii)} $\varepsilon>|h|$: 
the two quantum-well states involved in the dot solution
belong to different branches, $\varepsilon_{+}$ and $\varepsilon_{-}$, 
with real values of $K$, noted $\Ka$ and $\Kb$:
\begin{equation}\label{eq:state_caseii}
\Psi_m(\rho,\varphi) = c_\mathrm{a} \begin{pmatrix}
                     J_{m}(\Ka \rho)e^{im\varphi}\\ 
                     d_{+}(\Ka) J_{m+1}(\Ka \rho)e^{i(m+1)\varphi}
                   \end{pmatrix} + 
               c_\mathrm{b} \begin{pmatrix}
                     J_{m}(\Kb \rho) e^{im\varphi}\\  
                     d_{-}(\Kb) J_{m+1}(\Kb \rho)e^{i(m+1)\varphi}
                   \end{pmatrix},
\end{equation}
The boundary condition leads to
\begin{equation}
  J_{m}(\Ka) \, d_-(\Kb) J_{m+1}(\Kb)-J_{m}(\Kb) \, d_+(\Ka) J_{m+1}(\Ka) = 0 \,.
\label{eq:quantization_caseii}
\end{equation}

\noindent
\textit{Case iii)} $-|h|<\varepsilon<|h|$: 
one imaginary value of $K$, $\Ka \equiv i\kappa_\mathrm{a} $, and a real
value $\Kb$ are involved.
The energy associated to $\Ka$ is 
\begin{equation}
  \varepsilon_{\pm,\mathrm{a}} = -\kappa_\mathrm{a}^2 \pm 
                         \sqrt{-\kappa_\mathrm{a}^2\left(\alpha_n'+
                         \gamma'\kappa_\mathrm{a}^2\right)^2+h^2},
\end{equation}
and the coefficient for the wave function
\begin{equation}
d_{\pm}(\Ka) =  (-i) \frac{\varepsilon_{\pm,\mathrm{a}} + \kappa_\mathrm{a}^2 - h }
                 {\alpha_n' \kappa_\mathrm{a} + \gamma' \kappa_\mathrm{a}^3} 
       \equiv -i\delta_{\pm}(\kappa_\mathrm{a})\,.
\end{equation}
With $J_{m}(i \kappa \rho) = i^m I_m(\kappa \rho)$, the quantum-dot wave function 
is then written as 
\begin{equation}\label{eq:state_caseiii}
\Psi_m(\rho,\varphi) = i^m  c_\mathrm{a} \begin{pmatrix}
                       I_{m}(\kappa_\mathrm{a} \rho)e^{im\varphi} \\ 
                       \delta_{\pm}(\kappa_\mathrm{a}) 
                       I_{m+1}(\kappa_\mathrm{a} \rho)e^{i(m+1)\varphi}
                   \end{pmatrix} + 
               c_\mathrm{b} \begin{pmatrix}
                       J_{m}(\Kb \rho)e^{im\varphi} \\  
                       d_{-}(\Kb) J_{m+1}(\Kb \rho)e^{i(m+1)\varphi}
                   \end{pmatrix}\,.
\end{equation}
The boundary condition leads to
\begin{equation}
    I_{m}(\kappa_\mathrm{a}) d_{-}(\Kb) J_{m+1}(\Kb) - 
    J_{m}(\Kb) \delta_{\pm}(\kappa_\mathrm{a}) I_{m+1}(\kappa_\mathrm{a})= 0 \,.
\label{eq:quantization_caseiii}
\end{equation}
\end{widetext}

\begin{figure*}
\centerline{\includegraphics[width=\linewidth]{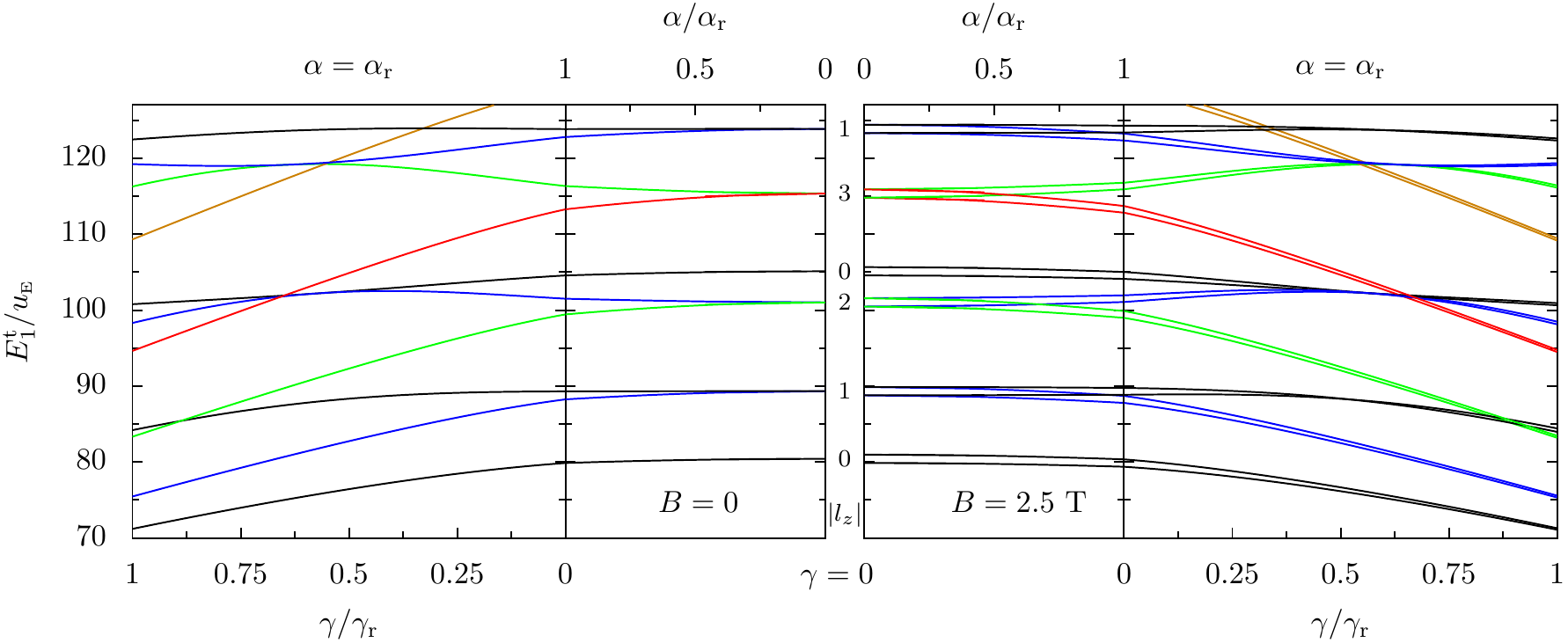}}
\caption{The discrete eigenenergies of a quantum dot with radius 
  $R=\unit[275]{\text{\AA}}$ and height $L=\unit[100]{\text{\AA}}$ for $n=1$,
  shown as a function of the spin-orbit coupling strengths
  $\alpha$ and $\gamma$, without magnetic field (left side) and 
  with $B=\unit[2.5]{\text{T}}$ (right side). 
  States with $|j_z|=1/2, 3/2, 5/2, 7/5$, and 9/5 are represented by 
  different black, blue, green, red, and orange lines, respectively.   
  In the central panels we keep $\gamma=0$ and vary $\alpha$
  from zero up to $\alpha_\mathrm{r} = \unit[0.571]{e\text{V\AA}}$,   
  reported in [\onlinecite{de-pry}]. 
  Conversely, in the following curves (outside panels) 
  $\alpha$ is fixed at $\alpha_\mathrm{r}$, 
  and $\gamma$ increases from zero to its final value of 
  $\gamma_\mathrm{r}= \unit[571.8]{e\text{V}\text{\AA}^3}$.
  In-between the panels, the values of the quantum numbers $l_z$ associated
  to the nearby states at zero SOC are indicated.
}
\label{fig:energyVsSoc}
\end{figure*}
Equations \eqref{eq:quantization_casei}, \eqref{eq:quantization_caseii},
and \eqref{eq:quantization_caseiii} express a root-finding problem, 
which we solve numerically. 
We find a family of solutions for each value of $m$ that correspond to the 
discretized energies of the quantum dot.
Moreover, the solutions give access to the wave numbers $\{\Ka,\Kb\}$ such 
that we can determine the coefficients $\{c_\mathrm{a},c_\mathrm{b}\}$ from 
the boundary condition and the normalization of the in-plane wave 
functions \eqref{eq:state_casei}, \eqref{eq:state_caseii}, and 
\eqref{eq:state_caseiii}.
All of these solutions carry a well-defined value of $j_z=m+1/2$, 
and in the absence of a magnetic field, the $j_z$ and $-j_z$ solutions are 
degenerate.

The results for the energy levels are presented in Fig.\ \ref{fig:energyVsSoc} 
as a function of the SOC coupling strength.
The states of different $|j_z|$ are shown with different colors.
To show the effect of the spin-orbit coupling, we start from the case
of vanishing SOC in the center of the figure and increase the SOC strength up 
to the predicted values $\alpha_\mathrm{r}$ and $\gamma_\mathrm{r}$ 
corresponding to the left and right edge of the figure. 

Without SOC ($\alpha=\gamma=0$, inner edges of the plot), the electronic states can be 
characterized by the orbital angular momentum $l_z$ along the $z$-axis and the 
spin $s=\pm 1/2$, in addition to the total angular momentum $j_z=l_z+s$. 
The values of $|l_z|$ corresponding to the states are indicated in the center 
of the figure. 
Without magnetic field (left side), the states characterized by
$(l_z,s)=(\pm |l_z|,\pm 1/2)$ are degenerate. In the presence of a 
magnetic field (right side), the Zeeman energy splits the levels corresponding 
different spin orientations.
In the presence of SOC, the orbital angular momentum and the spin get mixed, 
$l_z$ and $s$ cease to be good quantum numbers, and only the total angular
momentum quantum number $j_z$, shown by the different colors in 
Fig.\ \ref{fig:energyVsSoc} characterizes the states. 
It can be seen states corresponding to the same $|l_z|$ 
at zero SOC are split by the SOC according to the different values of $|j_z|$.   

In order to discriminate the effects of the different SOC terms, 
we increase the SOC in two steps.
We first consider the usual Rashba-like problem by setting $\gamma=0$ and 
varying the linear coupling strength $\alpha$ from zero up to 
$\alpha_\mathrm{r}=\unit[0.571]{e\text{V\AA}}$.
This situation is depicted in the inner part of Fig.\ \ref{fig:energyVsSoc}, 
where the left side corresponds to the case of zero magnetic field
and the right side to $B=\unit[2.5]{\text{T}}$.
The ensuing step is to fix $\alpha$ at $\alpha_\mathrm{r}$ and raise the 
value of $\gamma$ from zero to
$\gamma_\mathrm{r} = \unit[571.8]{e\text{V}\text{\AA}^3}$. 
The result is matched with the previous one and traced by the adjoining 
curves in the outer panels of the figure.
It must be noted that $\gamma$ determines not only the cubic-in-$k$ SOC 
coupling, but it also enters in the linear-in-$k$ coupling 
[cf.\ Eq.~\eqref{eq:alpha_redef}].
Consequently, at the end of each curve we find the energy of the quantum dot 
for the corresponding $\alpha_\mathrm{r}$ and $\gamma_\mathrm{r}$.

\begin{figure*}
\centerline{\includegraphics[width=\linewidth]{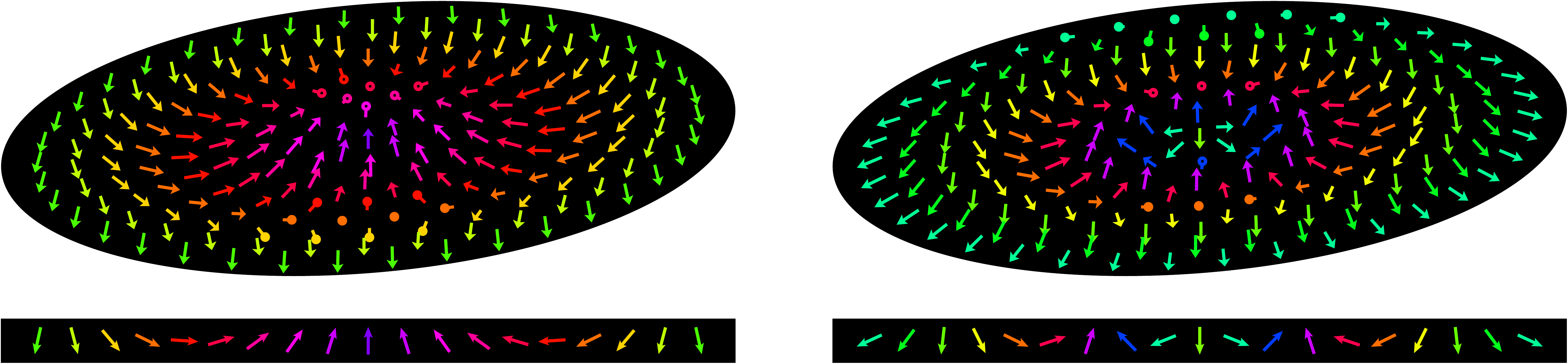}}
\caption{Spin textures in cylindrical quantum dots with 
  $L=\unit[100]{\text{\AA}}$ and $R=\unit[275]{\text{\AA}}$.
  Left and right panels show results for the lowest and the second lowest
  states with $|j_z|=1/2$. 
  The arrows and colors indicate the spin orientation as a function of the 
  position in the $xy$-plane.
  Below the disks, the same data are shown for a linear cut through the
  center of the sample.}
\label{fig:SpinTex}
\end{figure*}
Note the significant effect of $\gamma$ on the eigenenergies, that leads 
to much stronger energy changes than $\alpha$ alone.  
It brings, for example, the lowest pair of levels with $j_z =\pm 1/2$ 
(lowest black curves) down to energies that are below $E_1^z$.
Moreover, level crossings occur as a function of $\gamma$, changing 
the order of the states in energy with respect to the case of vanishing SOC.
This happens mainly for the lowest energy states of a given $|j_z|$ that are 
pulled down by the SOC below the higher energy states with lower values of 
$|j_z|$.
We remark that the full range of eigenenergies that we consider has not been 
explored in previous studies, and that we explicitly include allowed energy 
values that lie within the gap of the two-dimensional dispersion relation of 
Fig.\ \ref{fig:energyDisp}, that is $-|h|<\varepsilon_1 <|h|$.
It can also be observed that the Zeeman splitting shrinks as the SOC 
increases, while the spin mixture brought about by the latter increases 
accordingly. This indicates that the effective g-factor in quantum dots is 
affected by the SOC and depends on the geometry.


\subsection{Spin texture of the eigenstates}
\label{sec:spintexture}

We now investigate the properties of the quantum dot eigenstates. 
The spinor states of Eqs.\ \eqref{eq:state_casei}, \eqref{eq:state_caseii}, 
and \eqref{eq:state_caseiii} contain the information on the spin texture of the dot states. 
Without SOC and in the presence of a magnetic field, even a very weak one, 
the states are spin polarized, and the spin texture of the one-electron states 
is uniform throughout the dot. 
The appearance of a non-trivial spin texture is therefore a signature of 
the SOC, and can be seen as the degree of mixing of the two spin components 
in an eigenspinor. 
To obtain the spin texture corresponding to a state, we compute the
expectation value of the spin operator   
\begin{equation}\label{eq:spin_texture}
\langle \boldsymbol{\sigma} \rangle(\mathbf{r}) = 
           \Psi^{\dagger}(\mathbf{r})\boldsymbol{\sigma} \Psi(\mathbf{r})
\end{equation}
for each spatial point $\mathbf{r}$ inside the quantum dot. 
Because of the separability of the wave functions \eqref{eq:phigrande}, 
the spin orientation is independent of the longitudinal coordinate $z$.
Moreover, the rotational symmetry of the dots around the $z$-axis imposes 
that the resulting spin orientations present the same symmetry. 
Therefore, their projection on the $\hat{\varphi}$-direction vanishes, 
such that the local spin direction
\begin{equation}
\langle \boldsymbol{\sigma} \rangle(\mathbf{r})=
  \hat{r} \cos[\beta(r)] + \hat{z} \sin[\beta(r)] 
\end{equation}
has only radial and $z$-components.
The angle of the local spin orientation with respect to the $xy$-plane 
$\beta$ depends only on the radial coordinate $r$.
We construct the full eigenstate solution $\Psi$ with energy $\varepsilon$ as 
\eqref{eq:state_casei}, 
\eqref{eq:state_caseii}, and 
\eqref{eq:state_caseiii}, 
depending on the value of $\varepsilon$, with the corresponding $\Ka(\varepsilon)$ 
and $\Kb(\varepsilon)$ obtained from a numerical solution of the quantization
conditions 
\eqref{eq:quantization_casei}, 
\eqref{eq:quantization_caseii}, or 
\eqref{eq:quantization_caseiii}.

In Fig.~\ref{fig:SpinTex}, we present two examples of spin texture in 
cylindrical quantum dots of length $L=\unit[100]{\text{\AA}}$ and 
radius $R=\unit[275]{\text{\AA}}$, in the presence of the full linear 
and the cubic SOC terms with the coupling strengths $\alpha_\mathrm{r}$ 
and $\gamma_\mathrm{r}$ predicted in Ref.\ [\onlinecite{de-pry}]. 
The left panels show the dependence of the spin orientation on the position 
in the $xy$-plane for the lowest energy states that have $|j_z| = 1/2$.
This spin texture corresponds to one of the two sublevels in the lowest Zeeman doublet, 
shown in Fig.\ \ref{fig:energyVsSoc} by the two lowest black lines. The other of the 
sublevels, that are degenerate at $B=0$, has spin orientations with the sign of the 
$z$-component reversed. 
The right panels show the spin texture for the next higher levels that 
are characterized by $|j_z|=1/2$, corresponding to the second pair of 
levels (black lines starting at $|l_z|=1$ in Fig.\ \ref{fig:energyVsSoc}).


\subsection{Effective g-factor in quantum dots}
\label{sec:gfactor}

The effective g-factor is experimentally accessible, and it is thus a widely studied quantity. 
An example are the measurements of Ref.~[\onlinecite{bjo-fuh-han-lar-fro-sam}],
where the effective g-factor has been observed to depend on the dot size 
with absolute values that are reduced as compared to the bulk effective 
g-factor $g^{*}\approx -14.7$ (value from Ref.~[\onlinecite{de-pry-2007}]).
In the experiment, the effective g-factor is extracted from the linear term 
of the magnetic-field induced energy splitting
\begin{equation}
  \Delta E = \left| g_\mathrm{eff} \mu_{B} B \right|
\end{equation}
of two states that are characterized by the same $|j_z|$ and degenerate 
in the absence of a magnetic field. 
According to this definition, each quantum-dot state has its own effective 
g-factor, and we will focus on the effective g-factor of 
the ground state which is often the most relevant one. 
To calculate the effective g-factor we can use different approaches. 
The most direct way is to set the magnetic field strength to a small finite 
value, e.g.\ $B=\unit[0.1]{\mathrm{T}}$, and to calculate the difference between 
the two lowest dot energies, using the procedure of 
Sec.\ \ref{sec:dotenergies}.
Alternatively, in order to avoid the finite value of the 
magnetic field, we can express the effective g-factor as
\begin{equation}
  g_\mathrm{eff}=\frac{1}{\mu_{B}} \frac{\partial\Delta E}{\partial B}
                =g^*\frac{\partial\varepsilon}{\partial h}
\end{equation} 
in terms of the sensitivity $\partial\varepsilon/\partial h$ of the quantized dot 
energy levels with respect to the magnetic field, at $h=0$. 
To determine this derivative, we proceed as in the case of Rashba SOC 
treated in Ref.\ [\onlinecite{tsi-loz-gog}], and derive the quantization
conditions \eqref{eq:quantization_casei} and \eqref{eq:quantization_caseii} 
for negative and positive in-plane energy $\varepsilon$, respectively.
The resulting expression for the effective g-factor is 
\begin{widetext}
\begin{eqnarray}\label{eq:geff_analytic}
  g_\mathrm{eff}&=&-g^*\frac{\sgn(\varepsilon)u(\Ka)+u(\Kb)}{u(\Ka)u(\Kb)}
  \\ \nonumber
  &&\times\frac{J_m(\Ka)J_{m+1}(\Kb)}{\zeta(\Ka,\Kb)\left[2\Ka+\sgn(\varepsilon)
  u'(\Ka)\right]^{-1}
  +\sgn(\varepsilon)\zeta(\Kb,\Ka)\left[2\Kb-u'(\Kb)\right]^{-1}}\, ,
\end{eqnarray}
where we have defined the functions
\begin{equation}
  \zeta(\Ka,\Kb)=J_m(\Kb)J'_{m+1}(\Ka)+\sgn(\varepsilon)J'_m(\Ka)J_{m+1}(\Kb)
\end{equation}
\end{widetext}
and $u(K)=\alpha'K-\gamma'K^3$. We denote by $J'_m(K)$ and $u'(K)$ the 
derivatives of the functions $J_m$ and $u$ with respect to $K$.

\begin{figure}
\centerline{\includegraphics[width=\linewidth]{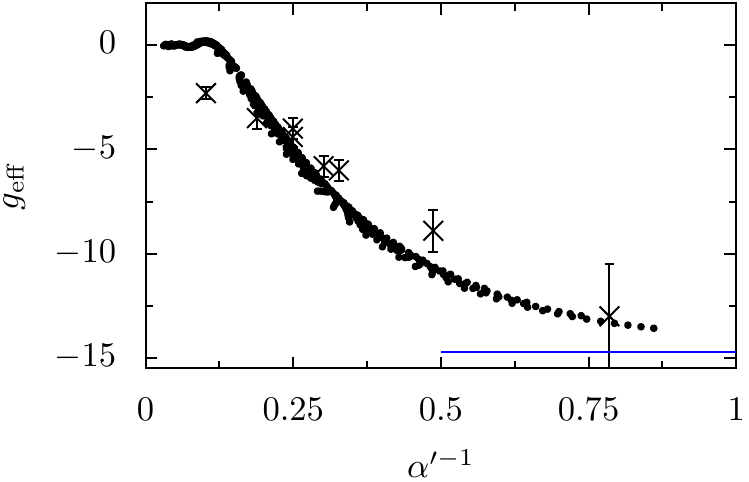}}
\caption{The calculated effective g-factors (full black circles) for cylindrical 
  quantum dots of different length $L$ and radius $R$, plotted versus
  $1/\alpha'$ defined in \eqref{eq:alpha_redef}, with the SOC parameters 
  from Ref.~[\onlinecite{de-pry}].
  The data points are for dots with radii $R$ from \unit[150]{\AA} to 
  \unit[500]{\AA}, and lengths values $L$ in the range between 
  \unit[50]{\AA} and \unit[200]{\AA}. The crosses represent experimental data 
  from Ref.\ [\onlinecite{bjo-fuh-han-lar-fro-sam}], obtained with a magnetic field 
  perpendicular to the symmetry axis of the quantum dot. 
  The blue horizontal line indicates the bulk effective 
  value $g^*\approx -14.7$.}
\label{fig:g}
\end{figure}

The expression of Eq.\ \eqref{eq:geff_analytic} is a generalization of the
result of Ref.\  [\onlinecite{tsi-loz-gog}], and it reduces to the result given 
in Eq.\ (13) of that paper in the case $\gamma=0$ of vanishing 
cubic-in-$\mathbf{k}$ SOC.  
In order to compute the effective g-factor using the analytic expression 
\eqref{eq:geff_analytic}, we first determine the eigenenergies and the
corresponding pair of wave-vectors $\Ka$ and $\Kb$ by solving numerically 
the quantization condition of Eqs.\ \eqref{eq:quantization_casei} and 
\eqref{eq:quantization_caseii} at $h=0$, and then evaluate 
\eqref{eq:geff_analytic} using the obtained values.   
In Fig.\ \ref{fig:g} we present our results for different dot dimensions 
with lengths ranging from \unit[50]{\AA} to \unit[200]{\AA} and radii 
from \unit[150]{\AA} to \unit[500]{\AA}.
We have checked that a direct numerical evaluation of the level splitting from 
numerically calculated energies at small values of magnetic field $B$ yields
the same results as Eq.\ \eqref{eq:geff_analytic}.  
In the figure, the numerical data for $g_\mathrm{eff}$ (black dots) is plotted 
as a function of the inverse effective dimensionless linear in-plane spin-orbit 
coupling $\alpha'^{-1}$ (see Eq.\ \eqref{eq:alpha_redef}). 
The data corresponding to different dot sizes approximately collapses on a 
single curve. While a plot as a function of $\alpha'$ shows the same data 
collapse, the presentation of Fig.\ \ref{fig:g} allows for greater clarity 
in the comparison with experiment.
Such a single-parameter scaling shows that the dependence of 
the ground state effective g-factor $g_\mathrm{eff}$ on $L$ and $R$ is, 
at least within the range of explored sizes, to a good approximation given by 
a function of $\alpha'$. Thus, the main mechanism giving 
rise to a size-dependence of the effective ground-state g-factor is the 
$L$-dependent renormalization of the effective linear coupling strength 
$\alpha'$ by the cubic SOC $\gamma$, and its scaling with $R$. 

For a fixed value of $L$, the renormalized linear-in-$k$ coupling 
strength $\alpha'$ is proportional to $1/u_{\text{\tiny E}} R$. 
Since $u_{\text{\tiny E}}\propto R^{-2}$, we have $\alpha'\propto R$ such that the 
effective linear coupling decreases with decreasing $R$. It can be seen in 
Fig.\ \ref{fig:g} that the value of $g_\mathrm{eff}$ increases (in absolute value) 
towards the bulk effective g-factor $g^*$ (blue line) as $R$ and thus the effective 
coupling $\alpha'$ decreases.
An increase in $R$ leads to a larger $\alpha'$, and according to Fig.\ \ref{fig:energyVsSoc}, 
the Zeeman splitting of the levels decreases as the SOC increases. 
The consequence is that $|g_\mathrm{eff}|$ diminishes.
Conversely, for a given radius $R$, the increase in $L$ leads to a decrease of
the effective linear-in-$k$ coupling $\alpha'$, with the result of an approach 
of $g_\mathrm{eff}$ to $g^*$. 

The effective cubic-in-$k$ coupling 
$\gamma' \propto 1/u_{\text{\tiny E}} R^3 \propto 1/R$ increases when $R$ decreases, 
and a competition between $\alpha'$ and $\gamma'$ can be expected. However, the spectrum 
of the lowest energies is related to small values of $K$ and mainly dominated by the linear SOC 
(see Fig.~\ref{fig:energyDisp}), at least for the not too small values of $R$ that we consider. 
The scaling of the results with $\alpha'$ leads to the conclusion that the main effect of the 
cubic Dresselhaus coupling $\gamma$ is the renormalization of the effective linear-in-$k$ coupling,
and that the impact of the effective cubic-in-$k$ coupling strength $\gamma'$ seems to be of minor 
importance. However, the above arguments are relevant for the case under study of not too small $R$ 
and low-energy dot states. More important effects of the cubic-in-$k$ coupling $\gamma'$ can 
be expected for the g-factor of excited states and in dots with very small $R$.
  
Similarly to the results presented in 
Ref.~[\onlinecite{de-pry-2007}], where a zincblende Hamiltonian with 
adjustable parameters such as the energy band gap magnitude was used, 
we find negative values for the ground state g-factor of the dot. 
However, in our case small positive values do occur for short 
pillbox-shaped dots.
In general, and similarly to the theoretical results for 
Rashba SOC \cite{tsi-loz-gog} as well as the experimental values of 
Ref.~[\onlinecite{bjo-fuh-han-lar-fro-sam}] (crosses in Fig.\ \ref{fig:g}), 
our effective g-factors are of reduced absolute value as compared to the bulk 
effective g-factor $g^*$. 
While the qualitative behavior and size-dependence of our results 
are clearly consistent with the data of Ref.\ [\onlinecite{bjo-fuh-han-lar-fro-sam}], 
a direct quantitative comparison cannot be made since in the experiment the magnetic field direction 
is not aligned with the symmetry axis of the dots. Also, while the effective g-factor has been measured 
for very different values of $L$, only a small range of radii has been covered in 
Ref.\ [\onlinecite{bjo-fuh-han-lar-fro-sam}].


\subsection{Phonon-induced spin relaxation rate}
\label{sec:spinrelaxation}

An important quantity characterizing the usefulness of quantum dots for 
spintronics applications is the spin lifetime, which is generally limited by 
interactions with acoustic phonons. 
We consider here the most relevant situation for possible applications, 
namely, an electron initially in the higher sublevel $\vert i \rangle$ of the lowest 
Zeeman doublet which relaxes to the lower sublevel $\vert f \rangle$ due to the 
emission of a phonon.
The rate $\Gamma$ of this process can be calculated using Fermi's Golden 
Rule
\begin{equation}\label{eq:fermi}
  \Gamma = \frac{2\pi}{\hbar} \sum_{\bff{Q},\lambda} 
  \vert \bra{f} U_{\lambda}(\bff{Q}) \ket{i}\vert^{2} [n(Q)+1] 
  \delta(\Delta E - \hbar \omega_{\lambda}) \, ,
\end{equation}
where $\bff{Q}$ is the phonon momentum. 
The label $\lambda = \{\mathrm{l, t}\}$ refers to the longitudinal and 
the transverse modes, respectively, and $n(Q)$ is the Bose-Einstein phonon 
distribution with energy $\hbar \omega_{\lambda}= \hbar c_{\lambda}Q$ where 
$c_{\lambda}$ is the sound velocity of the corresponding mode. 
The energy difference between the two electronic states 
$\Delta E = E_i -E_f$ determines via the $\delta$-function 
the energy of the phonons involved in the relaxation process.
The potential $U_{\lambda}(\bff{Q})$ comprises both the deformation and 
the piezoelectric contributions\cite{yin,yin-wu,zook,kok-com-dic} for 
wurtzite lattice structures. 
For the longitudinal mode, we have
\begin{equation}
 U_\mathrm{l}(\bff{Q}) = 
 \left[ \Xi_\mathrm{l}(\textbf{Q}) + i\Delta_\mathrm{l}(\textbf{Q})\right]
 e^{i\textbf{Q} \cdot \textbf{r}}
\end{equation}
with $\Xi_{\mathrm{l}}(\textbf{Q})$ being the deformation potential given by
\begin{equation}\label{eq:defo}
  \Xi_\mathrm{l}(\textbf{Q}) = \Xi_{0} A_\mathrm{l} \sqrt{Q} \, ,
\end{equation}
where $\Xi_{0}$ is a bulk-phonon constant. 
The quantity 
$A_\lambda=\sqrt{\frac{\hbar}{2V\varrho c_\lambda}}$ 
contains the mass density $\varrho$ and the sample volume $V$. 
The deformation potential has the same form as in the case of a
zincblende structure.
The term $\Delta_\mathrm{l}(\textbf{Q})$ accounts for the piezoelectric
contribution and upon introducing spherical coordinates 
$(Q, \theta_\mathrm{p}, \varphi_\mathrm{p})$ 
for the phonon momentum, it reads 
\begin{equation}\label{eq:piezo_l}
\Delta_\mathrm{l}(\textbf{Q}) = A_\mathrm{l}\; \frac{1}{Q^{1/2}} \Delta_{0} 
                                \cos\theta_\mathrm{p}
                \left(h_{33}-h_{x} \sin^{2}\theta_\mathrm{p} \right) \, , 
\end{equation}
where $h_{x} = h_{33} - 2 h_{15} - h_{31}$. 
In general, $h_{ij}$ are bulk phonon constants and 
$\Delta_{0} = 4\pi e /\kappa$, where $\kappa$ is the dielectric constant 
and $e$ the electronic charge. 
We emphasize that $\theta_\mathrm{p}$ is the angle between $\textbf{Q}$ and 
the $z$-axis (defined as the c-axis of the wurtzite structure). 

The potential of the transverse phonon mode is given by 
\begin{equation}
U_\mathrm{t}(\bff{Q}) = \Delta_\mathrm{t}(\textbf{Q}) e^{i\textbf{Q} \cdot \textbf{r}}
\end{equation}
with 
\begin{equation}\label{eq:piezo_t}
  \Delta_\mathrm{t}(\textbf{Q}) = A_\mathrm{t} \frac{1}{Q^{1/2}}
  \Delta_{0} \sin\theta_\mathrm{p}
  \left( h_{15} + h_{x} \cos^{2}\theta_\mathrm{p} \right)\, .
\end{equation}
We emphasize that in wurtzite lattices the transverse piezoelectric
potential has only one term, while in zincblende lattices it has two.
The matrix element in Eq.\ \eqref{eq:fermi} can be factorized, and 
the rate can be written as
\begin{equation}\label{eq:fermi2}
\Gamma = \frac{2\pi}{\hbar} \sum_{\bff{Q},\lambda} 
\vert M_{\lambda}(\bff{Q})\vert^2 
\vert \bra{f} e^{i \bff{Q}\cdot\bff{r}}\ket{i}\vert^{2} 
n(Q) \delta(\Delta E - \hbar \omega_{\lambda}) \, ,
\end{equation}
where $M_\mathrm{l}= \Xi_\mathrm{l}(\textbf{Q}) + 
i\Delta_\mathrm{l}(\textbf{Q})$ and 
$M_\mathrm{t} = \Delta_\mathrm{t}(\textbf{Q})$.
We first note that the modulus of the momentum is fixed by the 
$\delta$-function.
Concerning the integral over the electronic coordinates, we remark that both 
the initial and the final states denoted by 
$\Phi_{nm}^{f(i)}(\rho,\phi,z)= \Psi_{nm}^{f(i)}(\rho,\phi)
\sqrt{2/L}\sin(n\pi z/L)$ 
have the same $z$-dependent factor.
Therefore, the integral corresponding to the matrix element in 
\eqref{eq:fermi2} can be further split into two parts by using cylindrical
coordinates, leading to
\begin{equation}
  \vert \bra{f} e^{i \bff{Q}\cdot\bff{r}}\ket{i}\vert^{2} = 
  \vert Z(\theta_\mathrm{p})\vert^2 \vert\Upsilon(\theta_\mathrm{p},
  \varphi_\mathrm{p})\vert^2 \, .
\end{equation}
The integral over $z$ can be performed analytically, yielding
\begin{eqnarray}
  \vert Z(\theta_\mathrm{p})\vert^2 = \frac{2 (2\pi n)^4 (1- \cos q_z)}
  {q_z^2 \left[ (2\pi n)^2 - q_{z}^2 \right]^2 } \, ,
\end{eqnarray}
where the definition $q_z = Q L \cos(\theta_\mathrm{p})$ has been used.
The other integral $\Upsilon(\theta_\mathrm{p},\varphi_\mathrm{p})$ reads
\begin{widetext}
\begin{eqnarray}\label{eq:int_rho_phi}
  \Upsilon(\theta_\mathrm{p},\varphi_\mathrm{p}) &=& 
    \int_{0}^{1} \mathrm{d}\rho \, \rho
    \left[ \left(c_\mathrm{a}^{f} J_{m_f}(\Ka^f \rho) + 
                 c_\mathrm{b}^f J_{m_f}(\Kb^f\rho)\right)\,
    \left(c_\mathrm{a}^{i} J_{m_i}(\Ka^i \rho) + 
          c_\mathrm{b}^i J_{m_i}(\Kb^i\rho)\right) \right.\\ \nonumber
&&+ \left.
    \left( c_\mathrm{a}^{f} d_\mathrm{a}^f J_{m_f + 1 }(\Ka^f \rho) 
    + c_\mathrm{b}^f d_\mathrm{b}^f J_{m_f + 1}(\Kb^f\rho) \right)
    \left( c_\mathrm{a}^{i} d_\mathrm{a}^i J_{m_i+1}(\Ka^i \rho) 
    + c_\mathrm{b}^i d_\mathrm{b}^i J_{m_i+1}(\Kb^i\rho) \right) \right]  
    \\ \nonumber
&&\times \int_{0}^{2\pi} \mathrm{d}\varphi\, 
                         \exp\left[i(m_i-m_f)\varphi\right]\;
                         \exp\left[i Q \sin\theta_\mathrm{p}
                  \cos(\varphi_\mathrm{p} -\varphi)\rho R\right] \, .
\end{eqnarray}
The integral over $\varphi$ can be easily performed by applying the 
Jacobi-Anger relation
\begin{equation}\label{eq:jacobi}
  \exp\left[i x \cos\varphi \right] = 
  \sum_{l=-\infty}^{\infty} i^{l} J_{l} (x) \exp\left[i l\varphi\right] \, .
\end{equation}
Upon replacing Eq.~\eqref{eq:jacobi} in Eq.~\eqref{eq:int_rho_phi} and 
carrying out the integration over $\varphi$, terms for all values of $l$ 
vanish except the one with $l= m_f - m_i$. 
The integral then results in
\begin{equation}\label{eq:int_phi}
  \int_{0}^{2\pi} d\varphi\, e^{i(m_i-m_f)\varphi}
  e^{i Q \sin\theta_\mathrm{p}\,\cos(\varphi_\mathrm{p} -\varphi)\rho R} = 
  2\pi\; e^{i(m_i-m_f)(\varphi_\mathrm{p} - 
  \pi/2)} J_{m_f-m_i}(\rho R Q \sin\theta_\mathrm{p}) \, .
\end{equation}
\end{widetext}
As it can be seen in Eq.~\eqref{eq:int_phi}, the complex exponential becomes 
a common factor in Eq.~\eqref{eq:int_rho_phi}, and leads to 
$\vert\Upsilon(\theta_\mathrm{p},\varphi_\mathrm{p})\vert^2 = 
f(\theta_\mathrm{p})$, 
which is not surprising, since the cylindrical symmetry is not broken by the phonon potential.

In addition to the determination of the values of $\Ka$, $\Kb$, and 
$\Delta E$ from numerically solving the quantization condition 
\eqref{eq:quantization_casei}, \eqref{eq:quantization_caseii},
and \eqref{eq:quantization_caseiii} as in Sec.\ \ref{sec:dotenergies},
the calculation of the spin relaxation rate still involves an integral over 
$\rho$, and a subsequent integration over $\theta_\mathrm{p}$ 
(since neither 
$\vert \Upsilon(\theta_\mathrm{p}, \varphi_\mathrm{p})\vert^{2}$ 
nor $\vert M_{\lambda}(\mathbf{Q})\vert^2$ depend on $\varphi_\mathrm{p}$, 
cf.\ Eqs.~\eqref{eq:defo},~\eqref{eq:piezo_l} and~\eqref{eq:piezo_t}), 
that we perform numerically as well.

\begin{figure}
\centerline{\includegraphics[width=\linewidth]{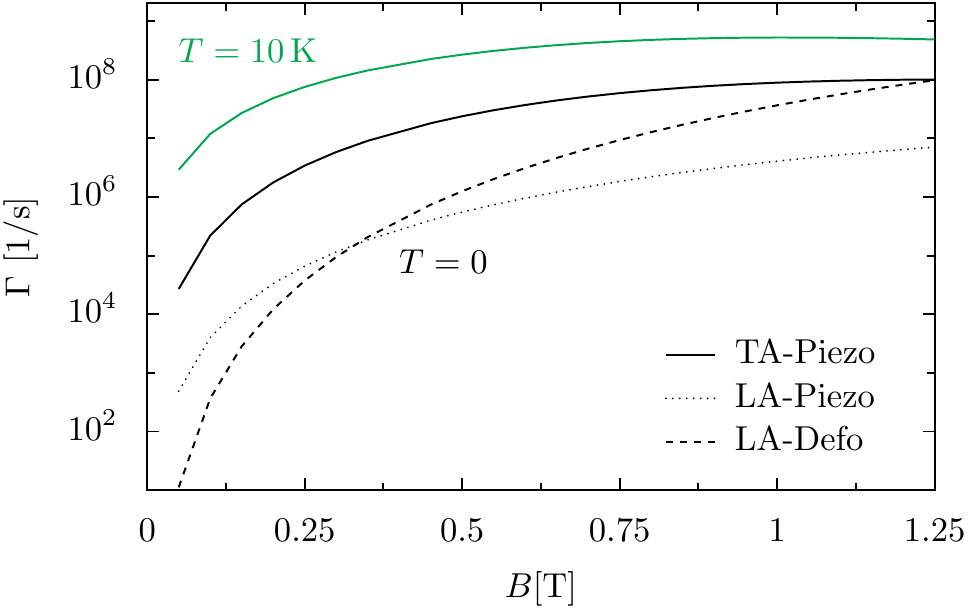}}
\caption{
  The calculated contributions of different acoustic-phonon potentials 
  to the spin relaxation rate as a function of magnetic field for cylindrical
  InAs quantum dots of length $L=\unit[100]{\text{\AA}}$ and radius 
  $R=\unit[275]{\text{\AA}}$, with wurtzite lattice structure. 
  The black curves correspond to the zero-temperature relaxation rate yielded 
  by the piezoelectric transverse (TA-Piezo; solid), the piezoelectric
  longitudinal (LA-Piezo; dotted), and the deformation (LA-Defo; dashed)   
  potentials. 
  The green line shows the spin relaxation rate due to TA-Piezo at finite
  temperature $T=\unit[10]{\text{K}}$.}
\label{fig:phonon}
\end{figure}
In the numerical evaluation of the relaxation rate, the parameters we use are 
$\varrho = \unit[5900]{\text{kg}/\text{m}^3}$, 
$c_\mathrm{l}=\unit[4410]{\text{m}/\text{s}}$,
$c_\mathrm{t}=\unit[2130]{\text{m}/\text{s}}$, 
$\kappa=15.15$, and 
$\Xi_{0} = \unit[5.8]{e\text{V}}$, 
all taken from Ref.~[\onlinecite{yin}].
As the piezoelectric bulk constants for InAs nanowires having wurtzite 
structure have not been obtained so far from microscopic calculations, 
we follow the standard prescription \cite{byk,yin-wu,yin,web-fuh-fas} 
of estimating them from the cubic structure by the use of the 
relations 
$h_{15} = h_{31} = (-1/\sqrt{3})h_{14}$ and 
$h_{33} = (2/\sqrt{3}) h_{14}$. 
For $h_{14}$ we use the value of the zincblende structure case 
($\unit[3.5 \cdot 10^8]{\text{V}\text{m}^{-1}}$~[\onlinecite{yin}]), 
assuming that at least the order of magnitude of that value should be 
correct for the wurtzite case. 
It can be seen from Eqs.\ \eqref{eq:piezo_l}, \eqref{eq:piezo_t}, and 
\eqref{eq:fermi2} that the relaxation rate due to the piezoelectric 
contributions is proportional to $h_{14}^2$ and thus not 
extremely sensitive to its precise value.

We present in Fig.~\ref{fig:phonon} the results for the spin relaxation rate 
in a cylindrical dot of length $L=\unit[100]{\text{\AA}}$ and radius 
$R=\unit[275]{\text{\AA}}$, as a function of magnetic field, assuming that 
the initial and final eigenstates in the relaxation process are the two 
lowest energy states (i.e the first Zeeman-split sublevels). 
The black curves show the zero-temperature contributions of the different 
phonon potentials separately. 
It can be seen that the transverse piezoelectric mode yields 
the dominant relaxation rate for magnetic field strengths 
below \unit[1.25]{T}. 
Moreover, the relaxation rate strongly increases with the magnetic 
field strength. 
This is due to the Zeeman splitting that makes phonons of higher energies 
relevant where the density of phonon states is increased. 
Quite long spin lifetimes, of the order of \unit[10]{ns}, occur for 
magnetic field strengths around \unit[1]{T}, and much longer lifetimes 
are obtained at weaker magnetic fields. 

The temperature dependence enters solely through the Bose-Einstein 
distribution in Eq.~\eqref{eq:fermi}, such that the increase of the 
spin relaxation rate with increasing temperature can be easily obtained. 
The result for the dominating TA-piezo mechanism at \unit[10]{K}
is shown in Fig.~\ref{fig:phonon} (green solid line). 

Though it is clearly dominated by the piezoelectric contribution, 
the order of magnitude of the deformation-coupling caused spin relaxation 
rates we find is consistent with the one of Ref.~[\onlinecite{yin}], 
where the singlet-triplet relaxation for an InAs nanowire-based quantum 
dot was calculated. 
However, in that work, only the deformation coupling was taken into 
account and assumed to dominate. 
The same assumption was made in Ref.~[\onlinecite{yin-wu}], where the 
electron spin relaxation in a similar quantum dot was calculated. 
In both references, the supposed dominance of the deformation over the
piezoelectric potential was justified by the fact that they considered 
small semiconductor nanostructures. 
As explained in Ref.~[\onlinecite{tak}], there is a competition between 
the two components that depends on the size of the nanostructure. 
For instance, the leading role of the piezoelectric coupling for weak 
magnetic fields has been reported \cite{rom-tam-ull} for 
quasi-one-dimensional ``cigar-like" quantum dots in GaAs nanowires 
with zincblende structure. 
For GaAs quantum dots, a crossing between the deformation and 
piezoelectric-induced rate as a function of the magnetic field was 
found in Ref.~[\onlinecite{des-ull}]. 
In our case this occurs as well, though for lower values of magnetic field 
than those observed in Ref.~[\onlinecite{des-ull}].
For InSb nanowires, numerical calculations show that the deformation potential 
dominates.\cite{rom-tam-ull,des-ull}
That domination has been assumed to be present in general, for all narrow-gap 
semiconductors.\cite{rom-mar-san} 
In contrast, in recent measurements on an InAs nanowire-based quantum 
dot,\cite{web-fuh-fas} the piezoelectric coupling was crucial for the 
determination of the phonon spectrum. 

We find that for InAs, which has a larger band gap than InSb, but smaller 
than GaAs, the spin relaxation rate is mainly driven by the (transverse)
piezoelectric phonon potential for magnetic fields below \unit[1.25]{T}. 
Beyond this value, the deformation seems to overcome the piezoelectric
contribution, but our theory does not allow us to treat stronger magnetic
fields.


\section{Concluding remarks}
\label{sec:conclusion}

Motivated by recent experiments on spin-orbit qubits
in InAs nanowire-based quantum dots,\cite{nad-fro-bak-kou} we have 
calculated the electronic structure of quantum wells and cylindrical 
quantum dots with wurtzite lattice structure, taking into account the 
linear and cubic Dresselhaus spin-orbit couplings and a weak applied 
magnetic field.
We found analytical solutions for the energy levels of both types of
structures. 
For quantum dots, we worked along the lines of the solutions previously 
found in the presence of the linear Rashba spin-orbit 
coupling.\cite{bul-sad,tsi-loz-gog}
Our obtained solution allowed us to explore the spin texture and the 
effective g-factor of the energy eigenstates, and, furthermore, we 
calculated the phonon-induced spin relaxation rate in the ground-state 
Zeeman doublet.

The energy dispersion of quantum wells shows a strong influence of
the cubic Dresselhaus term.
In quasi-two-dimensional structures, the cubic term leads 
to a thickness-dependent renormalization of the linear term which is
of the same order of magnitude as the bare term.
This renormalization produces a level crossing of all
the subbands (within the range of validity of perturbation theory)
that is being reported here for the first time.
Another level crossing, which is a consequence of the competition between 
the linear and the cubic terms, and which has been reported earlier in 
wurtzite-structure materials, is also obtained here.
Although our analytical results are valid for all wurtzite-lattice 
materials, for concreteness we presented full numerical results for InAs 
quantum wells only.

The quantum-dot eigenstates have been obtained as linear superpositions of
degenerate quantum-well states, and their associated energies calculated
as functions of the Dresselhaus coupling constants with and without
a relatively weak Zeeman term.
Again, a strong influence of the cubic term is observed in the energy levels.
The spin texture of the energy eigenstates is shown to be qualitatively
modified by the presence of the Dresselhaus spin-orbit coupling.
Further studies and applications of these states as qubits should take
into account this seldom discussed feature of the states.
Our analysis of the effective ground-state g-factor shows a remarkable 
scaling collapse when the data are plotted as a function of the effective 
linear in-plane spin-orbit coupling $\alpha'$, that contains a size-dependent 
renormalization from the cubic Dresselhaus coupling $\gamma$. 
The obtained results are consistent with existing experimental 
data \cite{bjo-fuh-han-lar-fro-sam} even though the magnetic field orientation 
is not the same. 
The scaling of the data indicates that other size-dependent mechanisms are of 
minor importance. However, the cubic in-plane coupling $\gamma'$ is expected 
to become more relevant in the case of very small radius $R$ and/or for the 
g-factor of higher excited states.   

Finally, we have calculated the acoustic-phonon-induced spin relaxation 
rate between the lowest Zeeman sublevels as a function of magnetic field.
The different rates arising from the longitudinal deformation, 
longitudinal piezoelectric, and transverse piezoelectric contributions
for wurtzite structures have been calculated and compared.
While our results for the spin relaxation rate due to the deformation 
mechanism are consistent with those of Ref.~[\onlinecite{yin}], 
we find that, in contrast to what is usually expected for small 
nanostructures,\cite{yin} the transverse piezoelectric phonon potential 
gives the dominant relaxation rate, at least for our case of cylindrical 
dots.

 
\acknowledgments
Acknowledgements: We thank Klaus Ensslin, Renaud Leturcq,
and Stevan Nadj-Perge for useful comments and Craig Pryor for helpful 
correspondence.
We gratefully acknowledge support from 
the ANR through grant ANR-08-BLAN-0030-02, 
the Coll\`{e}ge Doctoral Europ\'{e}en of Strasbourg, 
UBACYT 2011-2014, 
program ECOS Sud-Mincyt (Action A10E06), 
and the ITN European Project NanoCTM (Grant agreement 234970).


\end{document}